\documentclass[12pt]{article}
\usepackage{graphicx}
\usepackage{amsmath}
\usepackage{amssymb}
\usepackage{amsthm}
\usepackage{latexsym}

\begin{document}

\date{}

\title{\LARGE
{\bf Two-Dimensional Scaling Limits \\ via Marked Nonsimple Loops}
}

\author{
{\bf Federico Camia}
\thanks{E-mail: fede@few.vu.nl}\\
{\small \sl Department of Mathematics, Vrije Universiteit Amsterdam}\\
\and
{\bf Luiz Renato G. Fontes}
\thanks{E-mail: lrenato@ime.usp.br}\\
{\small \sl Instituto de Matem\'atica e Estat\'\i stica, Universidade de S\~ao Paulo}
\and
{\bf Charles M.~Newman}
\thanks{E-mail: newman@courant.nyu.edu}\\
{\small \sl Courant Inst.~of Mathematical Sciences,
New York University}
}


\maketitle

\begin{abstract}
We postulate the existence of a natural Poissonian marking of the double (touching) points
of $SLE_6$ and hence of the related continuum nonsimple loop process that describes
macroscopic cluster boundaries in 2D critical percolation.
We explain how these marked loops should yield continuum versions of near-critical percolation, dynamical percolation, minimal spanning trees and related plane filling curves,
and invasion percolation.
We show that this yields for some of the continuum objects a conformal
covariance property that generalizes the conformal invariance of critical systems.
It is an open problem to rigorously construct
the continuum objects and to prove that they
are indeed the scaling limits of the corresponding lattice objects.
\end{abstract}


\section{Introduction} \label{intro}

The rigorous geometric analysis of the continuum scaling limit of two-di\-men\-sional critical
percolation has made much progress in recent years.
This work has focused on triangular lattice site percolation, or equivalently random
(white-black) colorings of the faces of the hexagonal lattice, where the critical value
of the probability $p$ for a hexagon to be white is $p_c=1/2$.
A key breakthrough was the realization~\cite{schramm} (proved in~\cite{smirnov,cn3}) that
when the lattice spacing $\delta \to 0$, the limit of the $p=p_c$ lattice exploration path
must be chordal $SLE_6$.
This result was extended in~\cite{cn1,cn2} to obtain the ``full scaling limit" at $p=p_c$
of all macroscopic cluster boundaries as a certain $SLE_6$-based process of continuum
nonsimple loops in the plane.

In this paper, we propose an approach for obtaining the one-parameter family of
\emph{near-critical} scaling limits where $p=p_c+\lambda\delta^{\theta}$ as $\delta \to 0$
with $\lambda \in (-\infty,\infty)$ and $\theta$ chosen ($=3/4$ as in, e.g., \cite{bcks})
to get nontrivial $\lambda$-dependence in the limit (see also the discussions in~\cite{aizenman1,aizenman2}).
We do not present any theorems here but rather use heuristic arguments to provide what we believe
is the correct (or, at least, a correct) conceptual framework for not only scaling limits of
near-critical percolation but also, as we shall explain later, for related 2D scaling limits
such as of continuum percolation, Dynamical Percolation (DP), the Minimal Spanning Tree (MST)
and Invasion Percolation (IP).
Our purpose is to encourage work on the various problems associated with what we hope will be
the eventual rigorous implementation of this framework.

The framework relies on a postulated (natural) Poissonian ``marking" of the double points
(touching points) of $SLE_6$ and hence of the continuum nonsimple loops of the full scaling
limit at $p=p_c$.
Part of the motivation for such an approach is that an analogous one succeeded in the simpler
``Brownian Web" setting.
There, in place of $SLE_6$ double points, there were points in $(1+1)$-dimensional space-time
from which two distinct paths emerge~\cite{finr}.
In the Brownian Web setting, the Poissonian marking turned out to be rather straightforward
to implement.
An essential open problem, as discussed in Section~\ref{marking} after we review some properties
of $SLE_6$ and the critical loop process, is to find a rigorous implementation of the postulated
marking in the $SLE_6$ setting.
Key steps are to construct the correct intensity measure supported on $SLE_6$ double
points and show that it is the limit of the appropriate scaled counting measure on
the lattice.

In Sections~\ref{clusters}-\ref{invasion}, we then use the marked critical loops
to construct continuum versions of various lattice objects: near-critical white clusters
(Section~\ref{clusters}), near-critical loops (Section~\ref{loops}), (critical)
dynamical percolation (Section~\ref{dyn-perc}), near-critical exploration paths
(Section~\ref{exp-path}), minimal spanning trees and related lattice-filling curves
(Section~\ref{MST}), and invasion percolation (Section~\ref{invasion}).

\section{Marking the critical loops} \label{marking}

\begin{figure}[!ht]
\begin{center}
\includegraphics[width=8cm]{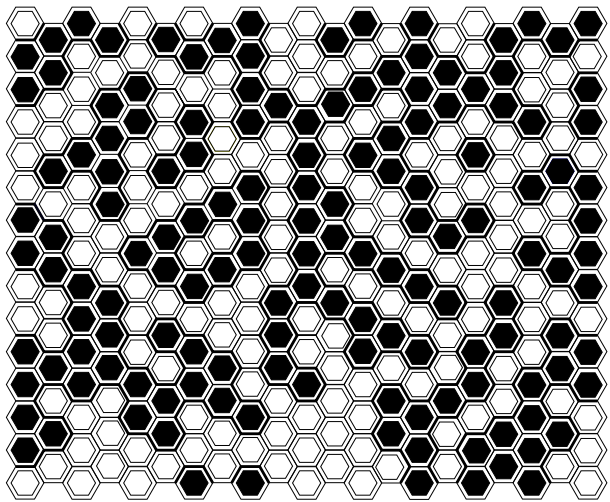}
\caption{Finite portion of a (site) percolation configuration
on the triangular lattice with each hexagon representing a
site assigned one of two colors.
In the critical percolation model, colors are assigned randomly
with equal probability.
The cluster boundaries are indicated by heavy lines; some small
loops appear, while other boundaries extend beyond the finite
region depicted.}
\label{percolation}
\end{center}
\end{figure}

In the critical ($p=1/2$) or near-critical ($p=1/2+\lambda\delta^{\theta}$) random
coloring of the faces
of the hexagonal lattice
(with lattice spacing $\delta$), we focus on the \emph{macroscopic} white and black
clusters, whose diameters are $\geq\varepsilon>0$ as $\delta \to 0$ (where $\varepsilon$
is arbitrary and will also be taken to $0$ eventually).
We consider the outer boundaries of these clusters as
a collection $X_{\epsilon}^{\delta}$ of simple closed curves in the plane
(modulo reparametrization -- see~\cite{ab,cn4,cn2} for technical details about this
and other issues) -- see Figure~\ref{percolation} -- which
we regard as oriented with a counterclockwise (respectively, clockwise)
orientation for the white (resp., black) clusters.
The limit in distribution,
as $\varepsilon, \delta \to 0$,
of this family of macroscopic cluster boundaries at $p=1/2$
is what we call the (critical) continuum nonsimple loop process or simply critical loop
process $X$ (with distribution $P$).
Here are a few of the almost sure properties of $X$ (or $P$).
\begin{itemize}
\item There are countably many loops, all continuous.
\item Loops do not cross themselves or each other.
\item Loops are nonsimple -- touching themselves and each other.
\item All touching points are double points -- there are no points in the plane touched
in total $3$ or more times (by $1,2$ or $3$ loops).
\item For any deterministic point $z$ and $0<r_1<r_2<\infty$, the number
${\cal N}(z,r_1,r_2)$
of loops surrounding $z$ that are contained in the annulus centered at $z$ with inner
(resp., outer) radius $r_1$ (resp., $r_2$) is finite but tends to infinity as $r_1 \to 0$
or $r_2 \to \infty$.
\item For any two loops $L, L'$, there is a finite sequence of loops
$L_1=L$, $L_2$, $\ldots,L_M=L'$ such that $L_i$
touches $L_{i+1}$ for each $i=1,\ldots,M-1$.
\item Every loop $L$ has a unique ``parent" loop $L^*$ which surrounds
$L$ (with no intermediate loop surrounding $L$); $L^*$ has opposite orientation
to $L$; $L$ has infinitely many ``daughter" loops.
\end{itemize}
We also note that the distribution of the critical loop process has various conformal
invariance properties including invariance under translations, rotations and scalings
of the plane.

\begin{figure}[!ht]
\begin{center}
\includegraphics[width=6cm]{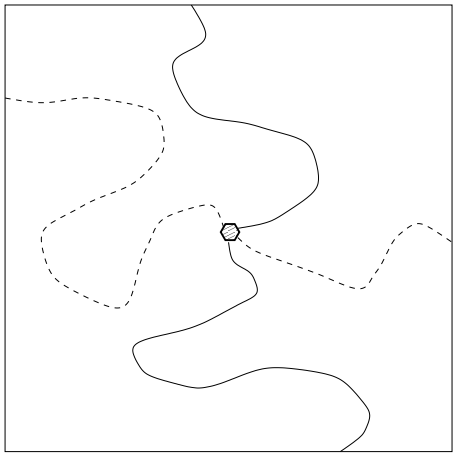}
\caption{Schematic diagram of a macroscopically pivotal hexagon at the center
of four macroscopic arms with alternating color. The full and dashed lines
represent paths of white and black hexagons respectively.}
\label{pivotal}
\end{center}
\end{figure}

The critical loop process may be constructed in the continuum by a procedure in which
each loop is a concatenation of two paths, each of which is distributed as (all or part
of) a chordal $SLE_6$ process -- the Schramm-Loewner evolution with parameter $6$.
This is of course based on the result~\cite{schramm,smirnov,cn3} that chordal $SLE_6$
is the scaling limit as $\delta \to 0$ of the critical percolation exploration path in
the closure $\overline D$ of a domain $D$, from $a$ to $b$ on the boundary $\partial D$.
The exploration path may be thought of as the (oriented, with white to its left) interface
between those white clusters which reach the counterclockwise $b$ to $a$ boundary and
those black clusters reaching the other portion of the boundary.
The first crucial idea of our marking process approach is that it is precisely the set
of double points (either where a single loop touches itself or where two loops touch)
that correspond to the $\delta \to 0$ limit of \emph{macroscopically pivotal} hexagons --
i.e., hexagons such that a color change (in either direction) has a macroscopic effect
on connectivity (or equivalently on loop structure).
In the lattice these are the hexagons that are at the center of four arms of alternating
color extending beyond macroscopic distance $\varepsilon$ from the pivotal hexagon itself
(see Figure~\ref{pivotal}).
In the limit of first $\delta \to 0$ and then $\varepsilon \to 0$, these give exactly
the set of all double points of the critical loop process.
For future use, we denote by $Y^{\delta}_{\varepsilon}$ the collection of random
points in the plane that are the centers of the $\varepsilon$-macroscopically pivotal
hexagons on the $\delta$-lattice.

The second crucial idea is that in near-critical percolation, as $\lambda$ varies, it is
only a special subset of the macroscopically pivotal hexagons that
actually have an effect, and these are the ones that need to be \emph{marked}.
To see this, it is best to use the standard coupling for different values of $p$ (and
hence different values of $\lambda$) provided by assigning i.i.d. random variables $U_h$,
uniformly distributed on $(0,1)$, to the hexagons (labelled by $h$) and (for given
$\delta$) say that a hexagon $h$ is $\lambda$-white if $U_h \leq 1/2+\lambda\delta^{3/4}$.
(As remarked in the Introduction, the value $\theta=3/4$ is chosen
in order to get nontrivial $\lambda$-dependence in the scaling limit -- see~\cite{bcks,cfn}).
Thus as $\lambda$ varies, a hexagon changes color when $\lambda$ crosses the value
$(U_h-1/2)\delta^{-3/4}$, and the only hexagons that change color as $\lambda$ varies in
$(-\infty,\infty)$ are those where $U_h-1/2$ is
$O(\delta^{3/4})$ as $\delta \to 0$.
The basic ansatz of our approach is that macroscopic changes in connectivity, loops etc.
in the scaling limit are those caused by macroscopically pivotal hexagons which change
color as $\lambda$ crosses $(U_h-1/2)\delta^{-3/4}$.
Our marking process marks each such site with the value where it changes
from black to white as $\lambda$ increases past that value.

It is known (see~\cite{sw,mz})
that with an $\varepsilon$-cutoff in the arm-length,
there are of order $\delta^{-3/4}$ macroscopically pivotal sites in any bounded region
of the plane and thus, as $\lambda$ varies in a bounded interval, there should be $O(1)$
such changing sites.
All this should be taken into account automatically by our postulated Poisson marking
process, as follows.

Let $P^{\delta}$ be the
probability
distribution of the collection
$X^{\delta}$
of directed cluster boundary loops on the $\delta$-lattice.
Conditional on a realization $x^{\delta}$ of
$X^{\delta}$,
we have denoted by
$Y^{\delta}_{\varepsilon}(x^{\delta})$ the collection of $\varepsilon$-macroscopically
pivotal hexagons on the $\delta$-lattice.
Let us now denote by $\nu^{\delta}_{\varepsilon}$ the (infinite) counting measure on
$Y^{\delta}_{\varepsilon}$, rescaled by a factor ${\delta}^{3/4}$, which assigns to a
subset $A$ of the plane the measure
\begin{equation}
\nu^{\delta}_{\varepsilon}(A)={\delta}^{3/4}\sum_{y \in A \cap Y^{\delta}_{\varepsilon}(x^{\delta})} 1.
\end{equation}
It has been proved~\cite{cn2} that $X^{\delta}$ converges in distribution, as $\delta \to 0$,
to an $X$ (with distribution $P$) given by (the directed version of) the $SLE_6$-based
continuum nonsimple loop process of~\cite{cn1,cn2}.
Two key open problems (which should perhaps be attacked in the opposite order) are (1)
to show that the pair $(X^{\delta},\nu^{\delta}_{\varepsilon})$ converges (jointly) in
distribution as $\delta \to 0$ to some $(X,\nu_{\varepsilon})$ with $\nu_{\varepsilon}$
(conditional on the realization of $X$) a locally finite measure supported on the double
points of $X$ and (2) to understand the nature of $\nu_{\varepsilon}$ (conditional on $X$)
in terms of $SLE_6$.
Note that because of the $\varepsilon$-cutoff, $\nu_{\varepsilon}$ is supported only
on either (a) double (touching) points of two loops with each having diameter
$\geq \varepsilon$ or (b) double points of single loops such that (as we explain more fully below in
Section~\ref{loops}) partitioning the single loop into two ``sub-loops" at the double
point yields both sub-loops with diameter $\geq \varepsilon$.

The intensity measure of the postulated Poissonian marking process, conditional on
a realization of the critical loop process $X$, should be the product measure
$\nu_{\varepsilon} \times d\lambda$ where $\nu_{\varepsilon}$ is the just-discussed
locally finite measure on double points of $X$ and $d\lambda$ is Lebesgue measure on
$(-\infty,\infty)$ for the assignment of a threshold value $\lambda$ to marked points.
Thus, there should be a (random) finite number of marked points located in bounded
regions of the plane with value $\lambda$ in bounded subsets of $(-\infty,\infty)$.
We expect this number to diverge as $\varepsilon \to 0$ (probably even if restricted
to those double points where another loop touches a single fixed loop of $X$).

\section{$\lambda_0$-connectivity} \label{clusters}

In order to discuss in the next section the near-critical loop process, depending on
$\lambda$, we need to discuss in this section the notion of $\lambda_0$-white (and
$\lambda_0$-black) regions and $\lambda_0$-white (and $\lambda_0$-black) curves in
the plane.
We begin by discussing the case $\lambda_0=0$ corresponding to the critical ($\lambda_0=0$)
loop process of~\cite{cn1,cn2}.

A $0$-white region (respectively, $0$-black region) is the union of the image of a
counterclockwise $0$-loop $L$ (i.e., one in the critical loop process) with its
``interior" less the union of the ``interiors" of all its (clockwise) ``daughter" loops.
Since these loops are nonsimple, the meaning of ``interior" needs to be explicated -- we
define it as follows.

For a counterclockwise (resp., clockwise) loop $L$ represented by the continuous curve
$\gamma:[0,1] \longrightarrow {\mathbb R}^2$ (with $\gamma(0)=\gamma(1)$), the open set
${\mathbb R}^2 \setminus \gamma([0,1])$ is the union of its countably many open, connected
components $\Gamma_0,\Gamma_1, \Gamma_2, \ldots$ with $\Gamma_0$ unbounded and $\Gamma_i$
bounded for $i \geq 1$, with each winding number ${\cal W}(\Gamma_i)$ of $\gamma$ about
$\Gamma_i$ (i.e., about any point in $\Gamma_i$)
for $i \geq 0$ equal to $0$ or $+1$ (resp., $0$ or $-1$).
The interior $\text{int}(L)$ of a counterclockwise (resp., clockwise) $L$ is
defined as the union of all $\Gamma_i$ with $i \geq 1$ and ${\cal W}(\Gamma_i)=+1$
(resp., ${\cal W}(\Gamma_i)=-1$). The exterior of $L$ is defined as the union of
those $\Gamma_i$ (including $\Gamma_0$) with ${\cal W}(\Gamma_i)=0$.

We denote by $L^*$ the closure (as a subset of ${\mathbb R}^2$) of the open set
$\text{int}(L)$ and claim that almost surely $L^* \setminus \text{int}(L)$
is equal to $\text{tr}(L)$ -- the trace of $L$, i.e. $\gamma([0,1])$.
(This is not true in general -- a counterexample is a curve whose trace is the
union of two disjoint circles and a segment joining them, where the points on the segment are
double points of the curve -- but in our case it follows from the fact that almost surely
the trace of a loop $L$ does not ``stick to itself." This, in turn, is a consequence of
the Hausdorff dimension of the double points of the trace of an $SLE_6$ curve.)
The $0$-white (resp., $0$-black) region $\tilde L$ defined by the counterclockwise
(resp., clockwise) loop $L$ is then
$L^* \setminus \cup_{L' \in {\cal D}(L)} \text{int}(L')$
where ${\cal D}(L)$ denotes the countable collection of daughter loops of $L$,
or equivalently, $\tilde L$ consists of the union of $\text{tr}(L)$ and all
the traces $\text{tr}(L')$ of the daughter loops of $L$ and all other points
$z$ in the plane that are ``surrounded" by $L$ but not ``surrounded" by any daughter
$L'$.

%

Now we can define a $\lambda_0$-white curve for $\lambda_0 \geq 0$
($\lambda_0$-black curves for $\lambda_0 \leq 0$ are defined similarly)
as a curve $\beta$ in
the plane whose image is a subset of the union of one or more $0$-white regions and
which is not allowed to {\it cross\/} $0$-loops $L$
(although it may {\it touch\/} them)
except at $\lambda$-marked sites with
$ 0 \leq \lambda \leq \lambda_0$. For
$\lambda_0 =0$, this means that the curve stays within a single
$0$-white region and does not cross any $0$-loops.

The $\lambda_0$-white curves can be used to describe the scaling limits of certain
connectivity probabilities for the one-parameter family of near-critical models.
For example, the scaling limit of the probability that two disjoint regions of the
plane, $D_1$ and $D_2$, are connected by a white path in the percolation model with
$p=1/2+\lambda_0\delta^{3/4}$ is given by the probability that there is a $\lambda_0$-white
curve from $D_1$ to $D_2$.
Analogously, the scaling limit of the probability that
in the $\delta$-lattice there is a white crossing of the
annulus centered at $z$ with inner radius $r_1$ and outer radius $r_2$ is the probability
that there is a $\lambda_0$-white curve crossing the same annulus.

\section{$\lambda_0$-loop processes} \label{loops}

\subsection{Construction} \label{construction}

In the previous section, we gave a preliminary discussion
of $\lambda_0$-connectivity. Here we focus
on the related notion of $\lambda_0$-loops, i.e., the scaling limit as $\delta \to 0$
of the collection of all boundary loops when $p=1/2+\lambda_0\delta^{3/4}$.
By analogy with the critical ($\lambda_0=0$) case and the definition of
$0$-white curves there,
the guiding idea is to define $\lambda_0$-loops
in such a way that $\lambda_0$-white curves
\emph{do not cross} any $\lambda_0$-loop.
In order to ensure this, one needs to merge together various $0$-loops and split other
$0$-loops, where the merging and splitting takes place at marked double points and the
decision to merge or split depends on the value associated to the mark and on $\lambda_0$.
The result of all this merging and splitting will be the collection of $\lambda_0$-loops.

\begin{figure}[!ht]
\begin{center}
\includegraphics[width=8cm]{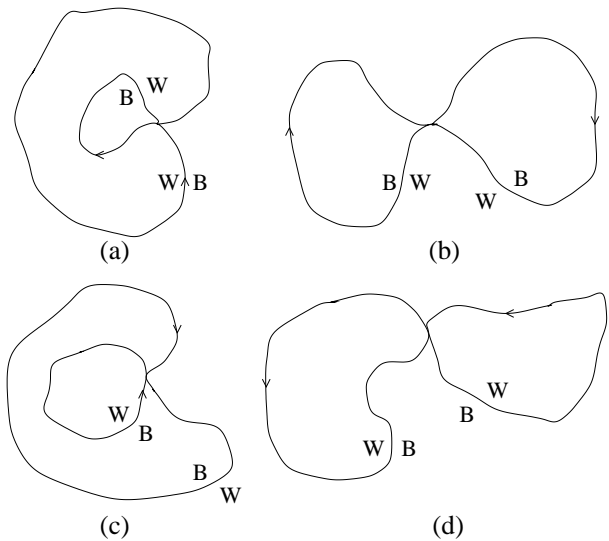}
\caption{Schematic diagram representing the different types of splitting and merging
caused by a black to white flip.
The arrows indicate the orientations of the loops, determined by having white (W)
on the left and black (B) on the right.}
\label{split-merg}
\end{center}
\end{figure}

For $\lambda_0>0$, the splitting of a $0$-loop (resp., merging of two $0$-loops),
caused by a black to white flip, takes place at a marked double point of the loop
(resp., where the two loops touch) with $\lambda \in [0, \lambda_0]$.
There are two types of splitting and two types of merging
(see Figure~\ref{split-merg}):
\begin{itemize}
\item[(a)] the splitting of a counterclockwise loop into an outer counterclockwise
and an interior clockwise loop,
\item[(b)] the splitting of a clockwise loop into two adjacent clockwise loops,
\item[(c)] the merging of a counterclockwise loop with the smallest clockwise loop
that contains it into a clockwise loop, and
\item[(d)] the merging of two adjacent counterclockwise loops into a counterclockwise
loop.
\end{itemize}
The case $\lambda_0<0$ is exactly symmetric to the
$\lambda_0>0$ case but
with splittings and mergings caused by white to black flips.

We point out that things are more complex than they may first appear, based on the
previous discussion.
This is because the critical loop process is scale invariant and each configuration
contains infinitely many loops at all scales, which implies that in implementing the
merging/splitting, one needs in principle both a small scale $\varepsilon$-cutoff and
a large scale $N$-cutoff.
This means that the merging/splitting should first be done only for loops touching a
square centered at the origin of side length $N$ and takes place only if both loops
involved in the merging or both loops resulting from the splitting have diameter larger
than $\varepsilon$.
For every $0<\varepsilon<N<\infty$, this ensures that the number of merging and splitting
operations is finite.
At a later stage, the limits $N \to \infty$ and $\varepsilon \to 0$ should be taken.

\subsection{Connectivity Probabilities}

The $\lambda_0$-loops can be used to describe the scaling limits of certain
connectivity probabilities for the one-parameter family of near-critical models.
For example, the scaling limit of the probability that two disjoint regions of the
plane, $D_1$ and $D_2$, are connected by a monochromatic path in the percolation
model with $p=1/2+\lambda_0\delta^{3/4}$ is given by the probability that there is
no $\lambda_0$-loop with one region in its interior and the other in its exterior
(where the interior and exterior are defined for $\lambda_0$-loops as they were
defined for $0$-loops in Section~\ref{clusters}).

\begin{figure}[!ht]
\begin{center}
\includegraphics[width=8cm]{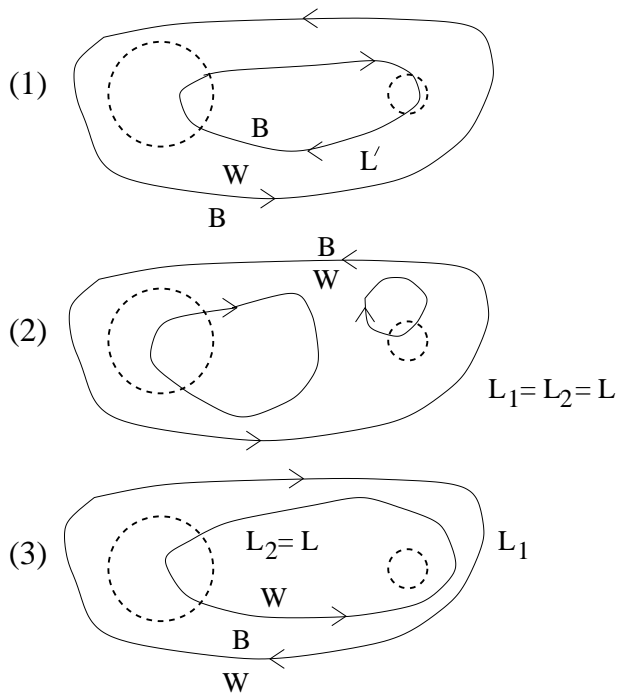}
\caption{Examples of monochromatic connections between the disc $D_1$ on the
left and the disc $D_2$ on the right. In (1), there are both white (W) and
black (B) connections; in (2) and (3), there is only a white connection.}
\label{connections}
\end{center}
\end{figure}

To determine whether there is a $\lambda_0$-white connecting curve
or a $\lambda_0$-black one or both, let $L_i$ for $i=1,2$ denote the
smallest $\lambda_0$-loop surrounding $D_i$.
In the situation we are considering where there is a monochromatic
connecting curve, there are three disjoint possibilities (see
Figure~\ref{connections}) -- either (1) there is a $\lambda_0$-loop $L'$
touching both $D_1$ and $D_2$, or (2) there is no such $L'$ and $L_1=L_2$,
in which case we define $L=L_1=L_2$, or (3) there is no such $L'$,
$L_1 \neq L_2$ and $L_2$ surrounds $L_1$ (resp., $L_1$ surrounds $L_2$),
in which case we define $L=L_1$ (resp., $L=L_2$).
Note that in case (3), $L_1$ touches $D_2$ (resp., $L_2$ touches $D_1$).
In case (1), there are both white and black connecting paths; in
cases (2) and (3), there is a white but no black connecting path
if $L$ is oriented counterclockwise, and otherwise there is a black
but no white connecting path.

\subsection{Conformal Covariance} \label{conf-cov}

The critical loop process has certain conformal invariance properties~\cite{cn2},
but one should not expect the $\lambda_0$-loop process to be conformally
invariant.
In fact, for every $\lambda_0 \neq 0$, the process should possess translation and
rotation invariance, but not scale invariance (see, e.g., Section 5 of~\cite{cfn}).

>From the fact that the $\lambda_0$-loop process is the scaling limit of a
percolation model with density $p=1/2+\lambda_0 \delta^{3/4}$, it can be
seen that a global dilation by a factor $s$ would map the $\lambda_0$-process
to a ${\hat \lambda_0}$-process with ${\hat \lambda_0}=s^{-3/4} \lambda_0$.
We will call this property of the one-parameter family of loop processes
\emph{scale covariance}.

Consider now the restriction of a $\lambda_0$-loop process to the subdomian $D$
of the plane, and let us assume for simplicity that $D$ is a Jordan domain (i.e.,
a bounded, simply connected domain whose boundary is a simple loop).
This can be obtained by taking the scaling limit inside $D$ (with some care in the
choice of ``boundary conditions" -- we don't dwell on this issue here, since it is
not important for our discussion), and it admits a construction via marked double
points analogous to that of the $\lambda_0$-loop process in the plane (see
Section~\ref{construction}).

In the critical case ($\lambda_0=0$), if $f:D \to {\hat D}$ is a conformal map from $D$
to ${\hat D}$, the $0$-loop process inside $D$ is mapped by $f$ to the $0$-loop process
inside ${\hat D}$ (in distribution).
This expresses the conformal invariance of the critical loop process.
The previous considerations on the scale covariance of the one-parameter family
of loop processes suggest that, in general, a $\lambda_0$-loop process in $D$
will be mapped (in distribution) by $f$ to an inhomogeneous ${\hat \lambda_0}(w)$-loop
process in ${\hat D}$ where ${\hat \lambda_0}(w) = |f'(z)|^{-3/4} \lambda_0$ and $w=f(z)$.
We will call this property \emph{conformal covariance}.

We note that there is no difficulty in interpreting an inhomogeneous loop process
within our framework of marked loops.
Such a loop process should be simply interpreted as one in which the $\lambda_0$
value that determines which marked double points should be used in the splitting/merging
procedure (equivalently, can be crossed by a white/black curve) depends on the
spatial location of the marks.
Such inhomogeneous loop processes would be the scaling limits of spatially
inhomogeneous percolation models in which $p=1/2+\lambda \delta^{3/4}$ but with
$p$ and $\lambda$ depending on location in the $\delta$-lattice. Conformal covariance
is then not just a property of our original one-parameter family
of loop processes but rather of this extended family of inhomogeneous loop
processes.

Conformal covariance for the extended family means that a $\lambda(z)$-loop
process in $D$ is mapped by $f$ to a ${\hat \lambda}(w)$-loop process in
${\hat D}$ with $\hat\lambda(w)=|f'(f^{-1}(w))|^{-3/4} \lambda(f^{-1}(w))$.
For example,
in order to see a homogeneous
$\lambda_0$-loop process in ${\hat D}$, one needs to start
with an inhomogeneous $(|f'(z)|^{3/4} \lambda_0)$-loop process in $D$.

\subsection{Relation with the Renormalization Group} \label{rg}

In view of its scale covariance property, as pointed out by Aizenman~\cite{aizenman1},
the one-parameter family of homogeneous $\lambda_0$-loop processes bears an interesting
relation to the Renormalization Group picture.
Loosely speaking, the action of space dilations can be interpreted as a flow in the
space of $\lambda_0$-loop models, where a dilation by a factor $s$ corresponds to
a change of $\lambda_0$ by a factor $s^{-3/4}$.

No general formalism has been found for an exact representation of
renormalization group transformations as maps acting in some space,
but in this specific example, our framework allows us to make precise the
(otherwise vague) meaning of flow in the space of loop models.
This is so because all loop models are defined on the same probability space
(of critical loops and marks), and the flow simply corresponds to a change of
the threshold value $\lambda_0$ that determines which marked double points
should be used in the splitting/merging procedure (equivalently, can be crossed
by a white/black curve).
We stress that the key to this interpretation as a flow is the coupling between
different loop processes via the marking procedure of Section~\ref{marking}.

\section{Dynamical percolation} \label{dyn-perc}

Our framework allows for the description of a weak limit of dynamical
critical percolation~\cite{hps,ps,scst} under a suitable time scaling
(as well as the usual space scaling).

In the dynamical (site) critical percolation model (on the triangular
lattice), we have as initial condition a critical configuration of white
and black hexagons. The color of each site/hexagon evolves by flipping
successively to the other color according to a rate 1 Poisson process,
independent of those of the other sites.

We scale space, as previously, by $\delta$ and time by
$\delta^{-\theta}$ with $\theta = 3/4$ (so that
the rate of the Poisson process slows down by a factor
$\delta^{3/4}$ -- i.e., one unit of
rescaled time corresponds to $\delta^{3/4}$ units in the original
lattice model).
The same considerations as in Section~\ref{marking} lead to the ansatz
that, under this combined scaling, only the
flips of macroscopically pivotal hexagons are relevant. 
For an eventual rigorous construction, we may also
need to implement a cutoff procedure on small and large scales like
the one described in Section~\ref{construction}.

As discussed in Section~\ref{marking}, the set of
macroscopically pivotal hexagons between two almost-touching macroscopic
(large) loops (as well as those of a single almost-self-touching loop --
with the usual proviso that upon flipping any one of these hexagons, there
should result two (large) macroscopic loops) has cardinality of the order
of $\delta^{-3/4}$.
Thus the choice of $\theta = 3/4$ for the time variable
scaling implies that, in the limit
$\delta \to 0$, we should see the following dynamics on the set of
double points of the (critical) continuum loop process.
Consider the double points between two given loops (or
alternatively from a single loop) in
the initial configuration of (critical) loops.
The color flipping of the
(macroscopically pivotal) hexagons at the discrete level will
correspond in the scaling limit to a process of marking double points in the continuum.
The marking takes place according to a Poisson process in space-time
whose intensity measure is $\nu_{\varepsilon} \times dt$ with
$\nu_{\varepsilon}$
the measure on the set of double points in question
described in Section~\ref{marking}.
(Note that eventually, the limit $\varepsilon \to 0$ is taken.)
At an event of the Poisson process, a mark is assigned to a double point
at a certain time.
The effect is essentially the same as that of the marking in the static
model: when a mark appears at a double point of two touching loops, then those
loops merge at the marked point; if it is a single loop touching itself at
that point, then after the time of the mark we have two loops touching at that point
(see Section~\ref{loops} and Figure~\ref{split-merg} there).

We remark that this dynamics can be realized using the static
marking of the critical loops, as described in Section~\ref{marking}.
Consider the same set of double points of two given loops
or a single loop of the initial configuration of (critical) loops
as above, but this time we mark this set with the static $\lambda$
marks as in Section~\ref{marking}. The dynamics is a function of the
loops and static marks as follows. As in the previous description,
(dynamical) marks will appear at certain times at certain double points.
The set of those spatial points where this takes place is exactly the set
of double points with the static $\lambda$ marks. It remains to
say at which time a dynamical mark appears at a given such site:
the answer is simply that
it appears at a site with the static mark $\lambda_0$ at time
$|\lambda_0|$.

As noted in~\cite{schramm1},
the dynamical percolation scaling limit should not be expected to
be conformally invariant.
However, its construction in terms of the static marks shows that
it must possess the same
conformal covariance as the family of loop processes of Section~\ref{conf-cov},
with the role of $\lambda$ played by time $t$.
This means that if $f:D \to {\hat D}$ is a conformal map from $D$ to ${\hat D}$,
the dynamical percolation scaling limit inside $D$ at time $t_0$ is
mapped (in distribution) by $f$ to a dynamical percolation scaling
limit in ${\hat D}$ at a spatially dependent
time ${\hat t}_0(w)$ given by
${\hat t}_0(w)=|f'(z)|^{-3/4} t_0$, where $w=f(z)$.
This gives a ``relativistic" framework where time is not ``absolute,"
and suggests a positive answer to a question posed by Schramm
in~\cite{schramm1}.

\section{$\lambda_0$-exploration path} \label{exp-path}

Take the hexagonal lattice embedded in the plane so that one of
its axes of symmetry is parallel to the $y$-axis.
Consider the set of hexagons that intersect the upper half-plane
$\mathbb H$ and induce two infinite clusters by coloring the
hexagons touching the $x$-axis white if they lie to the left of a
given hexagon and black otherwise.
The other hexagons in $\mathbb H$ are colored white
or black independently, with equal probability.
With probability one, there are exactly two infinite clusters, one
white and one black.
The critical percolation exploration path in $\mathbb H$
is the interface separating the infinite
white cluster from the infinite black cluster.

In a seminal paper~\cite{schramm}, Schramm conjectured that in
the scaling limit this critical percolation exploration path
converges in distribution
to (the trace of) chordal $SLE_6$.
Shortly thereafter, a crucial part
of the connection between the exploration path and $SLE_6$
was made rigorous by Smirnov's proof~\cite{smirnov} of
convergence of crossing probabilities to Cardy's
formula~\cite{cardy}.
Smirnov~\cite{smirnov} also proposed a strategy to use his Cardy
formula results to prove Schramm's complete conjecture.
A detailed proof of the Schramm conjecture~\cite{cn4,cn3} followed a few
years later.

The (full) scaling limit of the critical \emph{half-plane}
percolation model with the white/black boundary conditions described
above
gives a \emph{half-plane} critical loop process together with a continuous,
infinite curve $\gamma$ distributed like the trace of chordal $SLE_6$
(in the upper half-plane).
$\gamma$ does not cross any loops but touches infinitely many of them
(and indeed is constructed by piecing together segments of the loops,
as discussed in~\cite{cn4,cn2}).


Consider now a percolation model in the upper half-plane where,
as before, the hexagons touching the $x$-axis are colored white if
they are to the left of a given hexagon and black otherwise, but
the remaining hexagons are colored white with probability
$p=1/2+\lambda_0\delta^{3/4}$ and black otherwise.
For $\delta$ fixed and $\lambda_0$ positive (resp., negative),
we are in the white (resp., black) supercritical phase.
However, the choice of boundary conditions implies that, together
with an (almost surely unique) infinite white (resp., black)
cluster, there is also an (almost surely unique) infinite black
(resp., white) cluster.
The $\lambda_0$-percolation exploration path
(for lattice spacing $\delta$) in the upper half-plane
is the interface separating the infinite
white cluster from the infinite black cluster of this percolation
model.

It is natural to conjecture that, as $\delta \to 0$, the scaling limit of the
discrete $\lambda_0$-exploration path converges in distribution to
a continuous path, and
further that this continuum $\lambda_0$-exploration path
can be obtained from the scaling limit of the critical exploration
path (i.e., chordal $SLE_6$) and the half-plane critical loops by means
of the marking of double points discussed in Section~\ref{marking},
together with a suitably adapted version of the splitting and merging
procedure described in Section~\ref{loops}.

The splitting and merging of loops is analogous to that described in
Section~\ref{loops}; the novelty here consists in the self-touching of the
exploration path and in the ``interaction" between the exploration path
and the loops it touches.
The resulting double points should be marked as described in Section~\ref{marking},
so that at a marked double point, a critical loop can merge into the exploration
path or a new $\lambda_0$-loop can be created by splitting
off from the exploration
path.

For $\lambda_0>0$, the mergings and splittings are caused by black to
white flips of pivotal hexagons, and in the scaling limit correspond to
the following two situations:
\begin{itemize}
\item a counterclockwise critical loop touching the critical exploration
path at a marked double point can merge with it, becoming part of the
$\lambda_0$-path, or
\item the critical exploration path can split at a marked (self-touching)
double point into a ``shorter" path and a clockwise $\lambda_0$-loop
touching each other at the marked double point.
\end{itemize}
The case $\lambda_0<0$ is of course symmetric to the one described
here but with splittings and mergings caused by white to black flips.

Once again, as already explained in Section~\ref{loops}, things are more
complex than they may appear from the above description, and one needs in
principle both a small scale $\varepsilon$-cutoff and a large scale
$N$-cutoff (see Section~\ref{construction} for a discussion of this point).

The scaling limit of the $\lambda_0$-exploration path for $\lambda_0 \neq 0$
should not be expected to be scale invariant.
However, its construction in terms of splitting and merging of loops
at marked double points as described above shows that it should possess
a scale covariance property (see Section~\ref{conf-cov}).

To further explore how the scaling limit of the $\lambda_0$-exploration
path should change under conformal transformations, let us consider the
$\lambda_0$-exploration path inside a Jordan domain.
Given a Jordan domain $D$ with two distinct points $a,b$ on its boundary
$\partial D$, one can define the $\lambda_0$-exploration path inside $D$
from $a$ to $b$ just like the $\lambda_0$-exploration path in $\mathbb H$
from $0$ to $\infty$.
More precisely, one has to choose appropriate boundary conditions (say,
white on the counterclockwise arc $\overline{ba}$ and black on the
counterclockwise arc $\overline{ab}$) and then consider the unique
interface between the white cluster touching the counterclockwise arc
$\overline{ba}$ and the black cluster touching the counterclockwise arc
$\overline{ab}$.

It follows from the considerations of Section~\ref{conf-cov}
that if $f:D \to {\hat D}$ is a conformal map from
$D$ to ${\hat D}$, the scaling limit of the $\lambda_0$-exploration path inside
$D$ is mapped by $f$ to a curve ${\hat \gamma}$ in ${\hat D}$
from ${\hat a}=f(a)$ to ${\hat b}=f(b)$
whose distribution can be obtained in the following way.
Consider the full scaling limit inside ${\hat D}$ with white/black boundary
conditions, corresponding to an $SLE_6$ path $\gamma$ from ${\hat a}$ to ${\hat b}$
and a countable number of critical loops.
Then ${\hat \gamma}$ is distributed like the path obtained from $\gamma$ after
applying to it the splitting/merging procedure described above with a
spatially dependent threshold value given at $w=f(z)$ by $|f'(z)|^{-3/4} \lambda_0$.

\section{Minimal Spanning Trees and plane-filling curves} \label{MST}

In this section, we propose a construction of the scaling limit of the
discrete minimal spanning tree (MST), using our framework of continuum
nonsimple loops and marked double points.
For an interesting discussion of universality properties of the MST in
two (and higher) dimensions, see~\cite{read}.
The discrete MST is most easily defined on the square lattice, so in
this section we focus on bond percolation on ${\mathbb Z}^2$.
For each nearest neighbor bond (or edge) $b$, let $U_b$ be a uniform $(0,1)$
random variable with the $U_b$'s independent.
This provides a standard coupling 
of bond percolation models for all values $p$ of the probability that
a bond $b$ is open by saying that $b$ is $p$-open if $U_b \leq p$.
One then defines the minimal spanning tree in, say, an $N \times N$ square
$\Lambda_N$ centered at the origin, with free boundary conditions, as the
spanning tree in $\delta{\mathbb Z}^2 \cap \Lambda_N$ with the minimum value
of $\sum_b U_b$ summed over $b$'s in the tree.
It is known, based on the relation to invasion percolation, that there is
with probability one a single limiting tree as $N \to \infty$
(see~\cite{ccn,alexander,ns}).
We will denote this tree on the $\delta$-lattice, $\delta{\mathbb Z}^2$,
by $T_{\delta}$.

The purpose of this section is to describe the putative scaling limit
(in distribution) of $T_{\delta}$ as $\delta \to 0$, in terms of the
critical $0$-loop process and our $\lambda$-marked double points.
Our description uses a minimax construction in the continuum which is
a natural analogue of a well-known one on the lattice (see, e.g.,
\cite{alexander} and references therein).
We ignore differences between bond percolation on the square lattice and
site percolation on the triangular lattice in the belief that they have
no effect on the continuum scaling limit; in particular we will use
``white" and ``open" interchangeably.

\begin{figure}[!ht]
\begin{center}
\includegraphics[width=8cm]{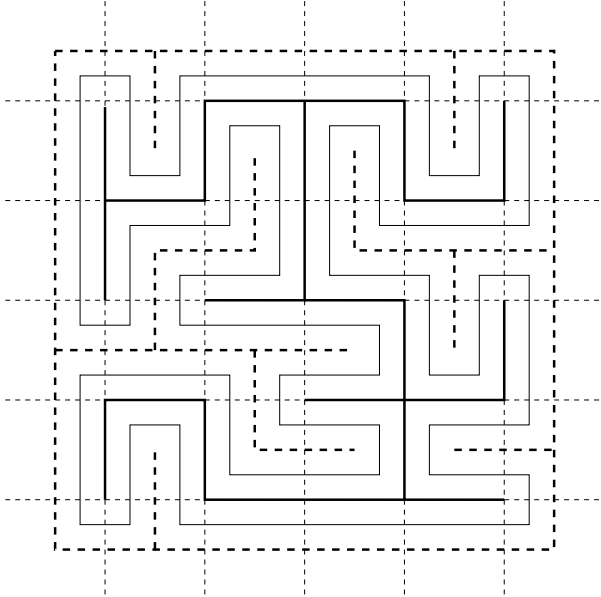}
\caption{Examples of a spanning tree (heavy line) with free boundary
conditions, its dual (heavy broken line) and the ``lattice filling"
curve between them.}
\label{mst}
\end{center}
\end{figure}

Once the (lattice) MST has been defined, a dual tree on the dual lattice arises naturally
as the unique spanning tree of the dual lattice which uses only those dual bonds that are
perpendicular to bonds that do not belong to the MST.
More precisely, given a minimal spanning tree $T_{\delta}^N$ in
$\delta{\mathbb Z}^2 \cap \Lambda_N$ with free boundary conditions, the dual tree is a
spanning tree $\tilde T_{\delta}^N$ on
(a portion of) the dual lattice $\delta[{\mathbb Z}^2+(1/2,1/2)]$ with
wired boundary conditions, i.e., with all the dual sites in the boundary
(of that portion of the dual lattice) identified and treated as a single site.
The edges of the dual tree $\tilde T_{\delta}^N$ cross only edges of ${\mathbb Z}^2$
that do not belong to $T_{\delta}^N$.
The dual tree is distributed like a minimal spanning tree (on the dual lattice)
with wired boundary conditions.
It can be shown that the $N \to \infty$ limits of the minimal spanning tree exist
for both free and wired boundary conditions
and moreover that the two limits coincide (see~\cite{ccn,alexander,ns}).
The spanning tree $T_{\delta}^N$ and its dual $\tilde T_{\delta}^N$ together define
a ``lattice-filling" curve $\pi_{\delta}^N$ as the curve that separates them (see
Figure~\ref{mst}).
With this choice of boundary conditions, $\pi_{\delta}^N$ is
actually a lattice-filling loop. In the limit $N \to \infty$, we drop the
$N$ superscript.

The MST $T_{\delta}$ can be decomposed into a forest
$\cal F$ ($= {\cal F}_{\delta}$) whose connected
components are the minimal
spanning trees of the individual critical ($p=1/2$) open clusters, such that
each ``cluster tree'' uses only bonds from a single open cluster
(including singleton site clusters when that site in the lattice
has all four of its touching edges closed), plus the set $\cal R$
of all the remaining bonds minimally connecting all those cluster
trees.
If we consider the individual open clusters as the vertices of a
graph $\mathbb G$ whose edges are the bonds connecting distinct open
clusters (and we may focus on those bonds with the smallest value
of $U_b$ for each adjacent pair of open clusters), we can define
a minimal spanning tree for $\mathbb G$.
Combining this minimal $\mathbb G$-spanning tree and the forest
$\cal F$ gives back the usual MST for $\delta {\mathbb Z}^2$.

In the scaling limit, we may consider the continuum MST as the limiting
set of paths  within the tree (see~\cite{abnw} for a general discussion
of continuum scaling limits of trees, which takes a somewhat
different point of view).
In order to define this tree, it is enough to describe the (unique, with
probability one) tree path between any two given deterministic points in
${\mathbb R}^2$.
However, it will be more convenient to describe the continuum tree path
between pairs of (non-deterministic) points, $z_1,z_2$, such that each
is contained in a continuum $0$-white region.
Since such points are dense in ${\mathbb R}^2$, one should obtain from
these the paths between all pairs of points, including deterministic ones.

For this purpose, we will use the idea of $\lambda$-connectivity
introduced in Section~\ref{clusters}.
Any two points $z_1,z_2$ contained in continuum $0$-white (open) regions
(of the critical model) should be $\lambda$-connected for some large enough
value $\lambda<\infty$.
To find the tree path between $z_1$ and $z_2$, we start decreasing
$\lambda$ from $+\infty$ until it reaches a value $\lambda_1$ below
which $z_1$ and $z_2$ are not $\lambda$-connected.
$\lambda_1$ is the smallest $\lambda$ for which $z_1$ and $z_2$ are
$\lambda$-connected and furthermore $\lambda_1$ should be the value of a
unique marked point $\zeta_1$ on all $\lambda_1$-paths from $z_1$ to $z_2$.
We then reduce $\lambda$ below $\lambda_1$ to a value $\lambda_2$ below
which either $z_1$ or $z_2$ is disconnected from $\zeta_1$.
This will give us a new marked point $\zeta_2$ labelled with a $\lambda$
equal to $\lambda_2$.
The procedure continues iteratively until the points $\zeta_i$ fill in a
continuous path between $z_1$ and $z_2$.

The procedure outlined above is the continuum version of a standard minimax
algorithm (see, e.g., \cite{alexander} and references therein) to construct
the minimal spanning tree on ${\mathbb Z}^2$ (using uniform $(0,1)$ bond
variables) where one looks at the minimum over all paths from $z_1$ to
$z_2$ of $\max_{b \in \text{path}}U_b$ to get a particular bond, and
then the procedure is repeated iteratively as above.

We note that the minimax value of $\lambda$ for the connection between points
in two different $0$-white (open) regions will be positive, while the minimax
value of $\lambda$ for the connection between points in the same $0$-white
(open) region will be negative.
The minimal spanning tree path between two points in the same continuum
$0$-white/open region is obtained by decreasing $\lambda$ from $0$ towards
$-\infty$, and the minimax points will be either double points of the
counterclockwise $0$-loop surrounding the cluster or points in the interior
of that counterclockwise $0$-loop where two clockwise daughter $0$-loops
touch each other or points where one such daughter loop touches the
mother $0$-loop.

We remark that we have presented our continuum minimax construction
of the continuum MST in a relatively simple version that does not use
any cutoffs (like those discussed in Section~\ref{construction}).
Even if such cutoffs turn out to be needed, the resulting construction
should still be feasible within our framework of loops and marked double
points.

The restrictions to the $0$-white/open regions of the continuum tree paths
correspond to the scaling limit of the forest $\cal F$ ($ = {\cal F}_{\delta}$),
while the minimal
$\mathbb G$-spanning tree converges in the scaling limit to a minimal
``cluster connecting tree," i.e., a minimal spanning tree in the ``graph"
whose vertices are the continuum $0$-white regions and whose edges
are the marked double points between them.

The scaling limit of the lattice-filling curve separating the MST and
its dual tree gives a continuous plane-filling curve closely
related to the $\lambda$-loops.
To see this, let us explain how to go from the plane-filling curve to
the $\lambda_0$-loops for some $\lambda_0>0$.
First of all, we give an orientation to the lattice-filling and
plane-filling curve so that the MST is to the left as one follows
the oriented curve.
Next, we note that the double points of the space-filling curve
are at marked points and are of two types, depending on whether they
occur at clockwise double points
or at counterclockwise double points -- i.e., ones where the
oriented loop formed by the curve between the first and second times
it touches the double point is clockwise or counterclockwise.
With this in mind, roughly speaking one proceeds as follows.
Starting from a marked counterclockwise double point with $\lambda>\lambda_0$
(resp., a clockwise double point with $\lambda<\lambda_0$), one begins by
following the plane-filling curve in the counterclockwise direction
(resp.,
the clockwise direction), taking shortcuts at every clockwise marked point
with $\lambda<\lambda_0$ and at every counterclockwise marked point with
$\lambda>\lambda_0$, until one returns to the starting marked point.
This gives a $\lambda_0$-loop.

To actually carry this out, one needs to take a double $(\varepsilon,K)$
cutoff, meaning that the marked points with $\lambda<-K$ or $\lambda>K$
are ignored and moreover a shortcut is taken only if the diameter of the
detour is at least $\varepsilon$.
The cutoff is removed by letting first $K \to \infty$ and then $\varepsilon \to 0$.

To conclude this section, we consider the scale and conformal invariance
properties of the continuum MST.
It is not known for sure whether or not the scaling limit of the MST is conformally
invariant and there does not seem to be general agreement on what to conjecture
(but a recent numerical study~\cite{wilson} suggests that the continuum MST is
\emph{not} conformally invariant).

Consider the scaling limit of the MST inside the Jordan domain $D$, and
let $f:D \to {\hat D}$ be a conformal map from $D$ to anther Jordan domain ${\hat D}$.
As we explained above, the lattice MST can be decomposed into a forest $\cal F$
($= {\cal F}_{\delta}$) and a minimal $\mathbb G$-spanning tree, where the latter
is conjectured to converge in the scaling limit to a minimal spanning tree in the
``graph" whose vertices are the continuum $0$-white regions and whose edges are the
(minimal) marked double points between them.
The construction of the continuum $0$-white regions from the (critical) $0$-loops
suggests that they should be conformally invariant.
Therefore the image of the set of continuum $0$-white regions of $D$ under
$f$ should be distributed like the set of continuum
$0$-white regions of ${\hat D}$.
However, from the considerations in Section~\ref{conf-cov}, we know that
if we transport to ${\hat D}$ the $\lambda$-marks associated to the double points
in $D$,
the distribution of the $\lambda$-marks that we obtain in ${\hat D}$ is not
the correct one, conditioned on the loops (and their double points); i.e.,
it is not the one that would have been obtained by carrying out the marking
procedure of Section~\ref{marking} in ${\hat D}$.
In fact, as explained in Section~\ref{conf-cov}, in order to get the correct
$\lambda_0$-loops in ${\hat D}$ from the loops obtained by mapping the critical
loops from $D$, one needs to rescale the $\lambda$-marks of double points
in $D$, typically in an inhomogeneous way.
This seems to suggest that the image under $f$ of the continuum MST in $D$
is typically not distributed like a continuum MST in ${\hat D}$, and therefore
that the scaling limit of the MST is not conformally invariant.

Note, however, that if the conformal transformation is a simple dilation,
the $\lambda$-marks need to be rescaled in a homogeneous way, so that the
distributions of the continuum $\mathbb G$-spanning trees obtained from
the marks before and after the rescaling should coincide, suggesting that
the minimal spanning tree is scale invariant (in distribution).

Another possible way to check the conformal invariance of the continuum MST
is by considering the distribution of the tree path between two given points,
$z_1,z_2 \in D$.
As explained above, this path can presumably be obtained via a minimax
algorithm that uses the marked double points in $D$ and their $\lambda$-marks.
The image under $f:D \to {\hat D}$ of the tree path between $z_1$ and $z_2$
is clearly the result of an identical minimax algorithm in ${\hat D}$ using
the $\lambda$-marks carried over from $D$.
On the other hand, the tree path in ${\hat D}$ between $f(z_1)$ and $f(z_2)$ is
obtained by the same minimax procedure using ``fresh" $\lambda$-marks,
assigned independently of the $\lambda$-marks in $D$.
But since a conformal map from $D$ to ${\hat D}$ typically changes the distribution
of the $\lambda$-marks in an inhomogeneous way, it seems likely that the
image under $f$ of the tree path in $D$ between $z_1$ and $z_2$ will not
be distributed like the tree path in ${\hat D}$ between $f(z_1)$ and $f(z_2)$.
This suggests once again that the scaling limit of the minimal spanning
tree is not conformally invariant.

\section{Invasion Percolation} \label{invasion}

We now turn to the scaling limit of (bond) invasion percolation
on ${\mathbb Z}^2$.
This also has a description in terms of the continuum marked loop
model as argued below.

The discrete model is as follows~\cite{wilwil,ccn}.
Each (nearest neighbor) bond of
${\mathbb Z}^2$ is assigned a positive continuous random variable
(say, uniformly distributed in $(0,1)$) -- called the resistance
of the bond -- independent from those of the other bonds.
Starting at a given site (say, the origin), sites are successively
invaded, one at a time, along bonds of least resistance among those
connecting previously invaded sites to uninvaded sites.
The invaded cluster (of the origin, at time infinity) is the
tree graph whose sites are all those eventually
invaded by this procedure and whose edges are all the bonds
which were invaded to reach these sites.

Let us color white the bonds with resistance smaller than $1/2$,
and color black the remaining bonds; let us also color the dual bonds
with the same color of the respective primal ones.
We then have that the set of white bonds form a critical percolation model.
Consider now the family of clusters of sites connected by the white bonds of
this percolation model (all of which are almost surely finite) and their
boundaries of black dual circuits.
It is clear from the rules of invasion that, before crossing any such
circuit (along any of its primal black bonds), all the sites in the
respective white cluster are invaded, after which the invasion proceeds
using the bonds of least resistance at the boundaries of the white clusters
in question.
After rescaling space by a small $\delta$, only macroscopic white clusters/black
circuits are relevant, and, following our ansatz and the rules of invasion,
only macroscopically pivotal bonds on black macroscopic circuits where two
macroscopic white clusters (almost) touch and which have resistance
$1/2+\lambda\delta^{3/4}$, $\lambda>0$, are relevant.

In the scaling limit, the following picture then emerges. Given the
system of critical counterclockwise loops along with their (positive)
$\lambda$-marks, suppose that a given set of $0$-white (critical) regions
(whose boundaries touch two by two) has been invaded thus far.
The invasion can be started from the smallest counterclockwise loop surrounding
the ball of radius $\varepsilon$ centered at the origin, where eventually
$\varepsilon \to 0$.)
The invasion thence proceeds by next incorporating the not currently invaded
$0$-white region whose outer boundary has the least $\lambda$ among all of the
marked double points between invaded and not yet invaded clusters.
The result of the invasion (after infinite time) can then be seen
as a graph whose sites are continuum $0$-white regions and whose edges
are certain of the marked double points between them.
Here, as before, we may need a cutoff procedure to insure that
the above least $\lambda$ value is realized.

To conclude this section, we briefly explain the connection between
minimal spanning trees and invasion percolation.
As argued in the previous section, the $\delta$-lattice MST can be decomposed
into a forest $\cal F$ of minimal spanning trees such that each tree uses only
bonds from a single open cluster, plus the set $\cal R$ of all the remaining
edges connecting those trees.
Each tree in $\cal F$ is obtained by doing invasion percolation ``with trapping"
inside an open cluster, that is, starting at a random site $x_0$ of the cluster,
the bond $b$ incident on $x_0$ with the smallest $U_b$ and the other site $x_1$
that $b$ is incident on are invaded, then the procedure is repeated by looking
at all the non-invaded bonds incident on $x_0$ and $x_1$, and so on, with the
proviso that a bond is not invaded if it is incident on two sites that are both
already invaded.
The procedure stops when all the sites of the clusters have been invaded, and
the above proviso ensures that the invaded bonds form a tree.
It is easy to see that the result is independent of the starting site $x_0$ and
that the tree obtained is the minimal spanning tree for the given cluster.
A similar invasion algorithm can be used to obtain the bonds in $\cal R$ by
considering the open clusters as the vertices of a graph $\mathbb G$ whose
edges are the ${\mathbb Z}^2$-paths between them, and doing invasion percolation
in $\mathbb G$.

\bigskip

\noindent {\bf Acknowledgements.}
The research of the authors was supported in part by the following sources:
for F.~C., a Marie Curie Intra-European Fellowship under contract
MEIF-CT-2003-500740 and a Veni grant of the Dutch Science Organization (NWO);
for L.~R.~G.~F., FAPESP project no.~04/07276-2 and CNPq projects
no.~307978/2004-4 and 475833/2003-1;
for C.~M.~N., grant DMS-01-04278 of the U.S. NSF.
This paper has benefitted from the hospitality shown to various of
the authors at a number of venues where the research and writing
took place, including the Courant Institute of Mathematical Sciences,
the Ninth Brazilian School of Probability at Maresias Beach, Instituto
de Matem\'atica e Estat\'\i stica - USP, and Vrije Universiteit Amsterdam.
The authors thank Marco Isopi and Jeff Steif for useful discussions.


\bigskip

\begin{thebibliography}{99}


\bibitem{aizenman1} M.~Aizenman,
Scaling limit for the incipient spanning clusters,
in {\em Mathematics of Multiscale Materials; the IMA Volumes in Mathematics
and its Applications} (K.~Golden, G.~Grimmett, R.~James, G.~Milton and
P.~Sen, eds.), Springer (1998).

\bibitem{aizenman2} M.~Aizenman,
Continuum limits for critical percolation and other stochastic geometric models,
in {\em XII Int.~Cong.~Math.~Phys. (ICMP '97, Brisbane)},
Internat.~Press, Cambridge, MA (1999).

\bibitem{ab} M.~Aizenman and A.~Burchard,
H\"{o}lder regularity and dimension bounds for random curves,
\emph{Duke~Math.~J.} {\bf 99}, 419--453 (1999).

\bibitem{abnw} M.~Aizenman, A.~Burchard, C.~M.~Newman, D.~Wilson,
Scaling Limits for Minimal and Random Spanning Trees in Two Dimensions,
\emph{Random~Struct.~Alg.} {\bf 15}, 319--367 (1999).

\bibitem{alexander} K.~S.~Alexander,
Percolation and Minimal Spanning Forests in Infinite Graphs,
\emph{Ann.~Probab.} {\bf 23}, 87--104 (1995).

\bibitem{bcks} C.~Borgs, J.~Chayes, H.~Kesten, J.~Spencer,
The Birth of the Infinite Cluster: Finite-Size Scaling in Percolation,
\emph{Comm.~Math.~Phys.} {\bf 224}, 153--204 (2001).

\bibitem{cfn} F.~Camia, L.~R.~G.~Fontes, C.~M.~Newman,
The Scaling Limit Geometry of Near-Critical 2D Percolation,
\emph{J.~Stat.~Phys.}, to appear (2006).

\bibitem{cn1} F.~Camia and C.~M.~Newman,
Continuum Nonsimple Loops and 2D Critical Percolation,
\emph{J.~Stat.~Phys.} {\bf 116}, 157--173 (2004).

\bibitem{cn4} F.~Camia and C.~M.~Newman,
The full scaling limit of two-dimensional critical percolation,
available at arXiv:math.PR/0504036 (2005), including a main part
corresponding to~\cite{cn2} and a long appendix roughly corresponding
to~\cite{cn3}.

\bibitem{cn2} F.~Camia and C.~M.~Newman,
The full scaling limit of two-dimensional critical percolation,
submitted (2005).

\bibitem{cn3} F.~Camia and C.~M.~Newman,
Critical Percolation Exploration Path and $SLE_6$: a Proof of Convergence,
in preparation (2006).


\bibitem{cardy} J.~L.~Cardy, Critical percolation in finite geometries,
\emph{J.~Phys.~A} {\bf 25}, L201--L206 (1992).

\bibitem{ccn} J.~T.~Chayes, L.~Chayes, C.~M.~Newman,
The Stochastic Geometry of Invasion Percolation,
\emph{Comm.~Math.~Phys.} {\bf 101}, 383--407 (1985).

\bibitem{finr} L.~R.~G.~Fontes, M.~Isopi, C.~M.~Newman, K.~Ravishankar,
Coarsening, Nucleation, and the Marked Brownian Web,
\emph{Ann.~Inst.~H.~Poincar\'e (B), Prob. and Stat.} {\bf 42}, 37--60 (2006).


\bibitem{hps} O.~H\"aggstr\"om, Y.~Peres, J.~E.~Steif,
Dynamical percolation, \emph{Ann.} \emph{Inst.~H.~Poincar\'e~Probab.~Statist.}
{\bf 33}, 497--528 (1997).

\bibitem{mz} G.~J.~Morrow and Y.~Zhang,
The sizes of the pioneering, lowest crossing and pivotal sites in
critical percolation on the triangular lattice,
\emph{Ann.~Appl.~Probab.} {\bf 15}, 1832--1886 (2005).

\bibitem{ns} C.~M.~Newman and D.~L.~Stein,
Ground State Structure in a Highly Disordered Spin Glass Model,
\emph{J.~Stat.~Phys.} {\bf 82}, 1113--1132 (1996).

\bibitem{ps} Y.~Peres, J.~E.~Steif,
The number of infinite clusters in dynamical percolation,
\emph{Probab.~Theory~Related~Fields} {\bf 111}, 141--165 (1998).

\bibitem{read} N.~Read, Minimal spanning trees and random resistor networks
in $d$ dimensions, \emph{Phys.~Rev.~E} {\bf 72}, 036114-1--036114-17 (2005).


\bibitem{schramm} O.~Schramm, Scaling limits of loop-erased random walks
and uniform spanning trees, \emph{Israel J. Math.} {\bf 118}, 221--288 (2000).

\bibitem{schramm1} O.~Schramm, Conformally invariant scaling limits:
an overview and a collection of problems, preprint math.PR/0602151 (2006).


\bibitem{scst} O.~Schramm, J.~E.~Steif,
Quantitative noise sensitivity and exceptional times for percolation,
preprint math.PR/0504586 (2005).


\bibitem{smirnov} S.~Smirnov, Critical
percolation in the plane: Conformal invariance,
Cardy's formula, scaling limits,
\emph{C.~R.~Acad.~Sci.~Paris} {\bf 333}, 239--244 (2001).

\bibitem{sw} S.~Smirnov and W.~Werner, Critical exponents for two-dimensional
percolation, \emph{Math.~Rev.~Lett.} {\bf 8}, 729--744 (2001).



\bibitem{wilwil} D.~Wilkinson and J.~F.~Willemsen, Invasion percolation:
A new form of percolation theory, \emph{J.~Phys.~A} {\bf 16}, 3365--3376 (1983).

\bibitem{wilson} D.~Wilson, private communication (2005).

\end{thebibliography}
\end{document}